\pdfoutput=1
\documentclass[conference]{IEEEtran}
\IEEEoverridecommandlockouts
\usepackage{cite}
\usepackage{amsmath,amssymb,amsfonts}
\usepackage{graphicx}
\usepackage{booktabs}
\usepackage{array}
\usepackage{tabularx}
\usepackage{textcomp}
\usepackage{url}
\usepackage[T1]{fontenc}
\usepackage[hidelinks]{hyperref}

\begin{document}

\title{Launch-Bound and Substitutable: Why Three Inference Optimizations Fail to Pay Off in Mixture-of-Experts Models}

\author{
\IEEEauthorblockN{Gokulakannan Sakthivel\IEEEauthorrefmark{1},
Jerry Wu\IEEEauthorrefmark{2},
Amogh Rajendra\IEEEauthorrefmark{1},
Giriprasad Radhakrishnan\IEEEauthorrefmark{1}}
\IEEEauthorblockA{\IEEEauthorrefmark{1}College of Computer, Mathematical, and Natural Sciences,
\IEEEauthorrefmark{2}Electrical and Computer Engineering\\
University of Maryland, College Park, MD, USA\\
\{gsakthiv, jerrywu, arajen15, rgpss02\}@umd.edu}
}

\maketitle

\begin{abstract}
Mixture-of-Experts (MoE) models route each token to a few of many expert networks, and that routing is data-dependent in a way standard inference optimizations do not expect. This paper measures what three of them actually deliver on OLMoE-1B-7B, DeepSeek-V2-Lite, and Qwen3-30B-A3B. Fused Triton kernels reach 5.6x to 9.0x in isolation but 0.999x end to end against a measured 1.07x ceiling, because the model spends its time waiting on roughly a thousand kernel launches per forward pass rather than on the arithmetic those kernels improve. INT4 quantization changes on average 0.53 of the eight selected experts per token position, yet replaying exactly those changed routes through full-precision weights reproduces only 2.7\% of the quality loss, which makes the experts substitutable rather than specialized. Removing all 23 graph breaks from PyTorch's compiler, the step prior work treats as the structural fix, makes the model three times slower. A fourth result ties the three together: leaving the routers in FP16 lowers drift by 20\% while raising loss, so routing fidelity and output quality are separable objectives. Every number recomputes from committed per-token route dumps.
\end{abstract}


\section{Introduction}

Large language models built on the Mixture-of-Experts (MoE) architecture have gained significant traction because they decouple model capacity from per-token compute cost. Rather than activating every parameter, a learned gating network selects a handful of expert feed-forward networks per token. This allows OLMoE \cite{ref1}, DeepSeek-V2 \cite{ref2}, and Qwen3 \cite{ref3} to hold billions of parameters while using only a fraction on any given forward pass.

Despite this advantage, optimizing MoE inference remains difficult. Unlike dense architectures, MoE inference relies on dynamic, data-dependent routing, where the subset of activated experts is determined by the embedding of each input token. That behaviour violates assumptions the usual optimizations are built on. Custom GPU kernels are written for fixed tensor shapes; quantization methods assume small weight perturbations produce small output changes; graph compilers such as torch.compile work best when the traced graph does not change between calls. The adaptive routing that makes MoE efficient is therefore also what complicates kernel optimization, model compression, and compiler acceleration.

Recent work has attacked the problem from the deployment side. MoE-squared plans collaborative inference across heterogeneous edge devices under latency and energy budgets \cite{ref16}. Liu et al. optimize how experts are distributed across serverless functions so that cold starts and cross-function traffic do not dominate \cite{ref17}. SP-MoE reorganizes the training pipeline so expert computation and communication overlap \cite{ref18}. All three take the per-device cost of a forward pass as given and schedule around it. We go the other way and ask what that per-device cost is actually made of, because if the single-device number is misattributed then every schedule built on top of it inherits the error.

This work investigates how the major inference optimizations interact with dynamic routing, and what the practical ceiling on each one really is. We evaluate system-level metrics (latency, throughput, memory) alongside the effect of each optimization on routing itself. Routing perturbation is quantified with four metrics: Routing Similarity, Jaccard Drift, Overlap@k, and Selection Shift. Crucially, we do not stop at correlating drift with quality. We intervene: the full-precision model is forced to follow the quantized model's expert choices, which separates what routing changes cost from what weight error costs. All experiments ran on NVIDIA A100 80GB SXM4 GPUs provisioned as serverless containers through Modal, one container per stage, so that no two measurements share a warmed device.

Three text-only models with clean FP16 baselines let us test whether the findings survive a change of routing configuration. OLMoE-1B-7B has 64 experts per layer with top-8 routing across 16 MoE layers, activating about 1B of 7B parameters. DeepSeek-V2-Lite uses 64 routed experts plus 2 shared, top-6, over 26 layers. Qwen3-30B-A3B is the widest at 128 experts, top-8, across 48 layers, and normalizes its top-k gate probabilities where OLMoE does not. Expert pool size, selection density, and gate normalization all differ, so a constraint that holds across all three is unlikely to be an artifact of one routing configuration.

Contributions: (1) A measured Amdahl ceiling of 1.07x for RMSNorm and router softmax, obtained by profiling wrapped modules rather than matching CUDA kernel names, and the finding that 5.6x to 9.0x isolated kernels still return 0.999x end to end because the model is launch-bound rather than arithmetic-bound. (2) A causal route-replay intervention showing that routing changes account for only 2.7\% of INT4's quality loss, with a control that verifies the intervention machinery is itself neutral. (3) Evidence that torch.compile's graph-break count is not a performance metric: eliminating all 23 breaks costs a factor of three in latency. (4) A router-exemption result in which drift and quality move in opposite directions, plus a top-k correction that makes drift comparable across models with different k and reverses the apparent cross-model ordering. (5) A reproducibility and execution layer built on Modal serverless A100 containers, where every stage commits its results to a persistent volume with a provenance manifest recording the git commit, resolved checkpoint revision, and package versions, and every published metric recomputes from raw per-token route dumps in the accompanying open-source repository with no GPU and no third-party packages.

\section{Literature Survey}

\subsection{MoE Architecture}

Sparse MoE models trace back to Shazeer et al. \cite{ref4}, who showed that model capacity can scale without a proportional increase in compute through a top-k gating function. The three models studied here inherit that design and differ mainly in how aggressively they sparsify. What matters for this paper is the consequence rather than the capacity: conditional, token-varying activation is the root cause of every optimization difficulty that follows.

\subsection{Custom GPU Kernels}

FlashAttention \cite{ref5} established that fusing an operation into a single SRAM-resident kernel removes bandwidth bottlenecks that no amount of arithmetic tuning would have reached. MegaBlocks \cite{ref7} carried the idea into MoE with block-sparse matrix multiplication. Our target is narrower. We fuse the dense operations that sit around the routing decision, RMSNorm \cite{ref6} and the gate softmax, because the dispatch itself has data-dependent output shapes that standard Triton tiling cannot express \cite{ref8}.

\subsection{Quantization}

GPTQ \cite{ref9} and bitsandbytes \cite{ref10} made post-training quantization practical for large language models, and SmoothQuant \cite{ref11} improved it further by moving the difficulty from activations into weights. The benefit is the reason quantization is the first optimization most deployments reach for: INT4 cuts weight memory roughly four-fold against FP16, which is what lets a 30B-parameter MoE model be served from a single 80GB device instead of two, and it raises the batch size that fits alongside the KV cache at a fixed memory budget. For MoE specifically there is a further attraction, in that the experts are the overwhelming majority of the parameters and are also the part touched least often per token, so compressing them looks close to free.

What has gone unexamined is whether that compression changes which experts get chosen. Quantization perturbs gate logits, and a top-k selection is a discrete function of those logits, so a perturbation smaller than the smallest logit margin changes nothing while one larger than it changes the expert set entirely. Frantar et al. \cite{ref9} report perplexity and do not examine selection stability; Dettmers et al. \cite{ref10} analyse outlier features in dense layers. We measure the routing topology change directly, and then test whether it matters.

\subsection{Graph Compilation}

torch.compile \cite{ref13} traces Python bytecode with TorchDynamo and lowers the result through TorchInductor, typically gaining 10 to 50\% on dense transformers. MoE routing produces graph breaks because conditionals on CUDA tensors and dynamic-shape operators cannot be traced statically. He et al. \cite{ref14} document conditional-flow breaks in serving workloads but do not characterize them inside MoE gating.

The relationship to this paper is direct, and it is the reason we re-measured rather than cited. The prior framing treats a graph break as a quantity to be minimized, which implies that a configuration reaching zero breaks should be the fastest available. Our compile benchmark tests that implication on the real checkpoint instead of assuming it, and finds it false by a factor of three. Any future work that proposes a custom lowering rule for MoE dispatch needs that result first, because it establishes that the break count is a traceability diagnostic and not a performance target.

\subsection{Why We Chose This Approach}

We study the three optimization families independently but under identical model configurations and hardware, so that results are directly comparable. Each sub-study uses a different measurement methodology: microbenchmark plus Amdahl analysis for kernels, routing-topology metrics plus a causal intervention for quantization, and graph tracing plus a latency sweep for compilation. The design also lets the sub-studies cross-check one another. The launch-bound scaling measured in Sub-study 1, where thirty-two times the tokens cost 1.55 times the time, independently explains why Sub-study 3 finds compilation unprofitable, even though the two measurements were taken with entirely different tools.

\section{Methodology}

\subsection{Models Under Study}

Three open MoE checkpoints were selected with deliberately different routing configurations. Table 1 sets out the comparison. Each model was loaded from a pinned revision at FP16 to establish the reference routing, and quantized variants were produced from that same checkpoint rather than from a separately published quantized release, so that drift is measured against a genuine baseline.

\begin{table*}[t]
\caption{Architecture comparison of the three models}
\label{tab:arch}
\centering
\begin{tabular}{@{}lccc@{}}
\toprule
\textbf{Property} & \textbf{OLMoE-1B-7B} & \textbf{DeepSeek-V2-Lite} & \textbf{Qwen3-30B-A3B} \\
\midrule
Total parameters & 7B & 16B & 30B \\
Active per token & 1B & 2.4B & 3B \\
Routed experts & 64 & 64 + 2 shared & 128 \\
top-k & 8 & 6 & 8 \\
MoE layers & 16 & 26 & 48 \\
Selection density & 12.5\% & 9.4\% & 6.3\% \\
Gate module type & nn.Linear & nn.Parameter & nn.Linear \\
norm\_topk\_prob & false & false & true \\
\bottomrule
\end{tabular}
\end{table*}

Two details in Table 1 turn out to matter later. The gate module type differs: OLMoE and Qwen expose the router as an nn.Linear, which bitsandbytes replaces, while DeepSeek-V2-Lite holds its gate as a raw nn.Parameter inside a custom MoEGate class and therefore stays FP16 under the same quantization call. DeepSeek's drift is consequently upstream perturbation only, where OLMoE's is that plus error in the gate weights themselves. The other detail is norm\_topk\_prob, which rescales the surviving gate probabilities after top-k and so changes how an expert swap propagates into the output.

Each model is quantized from its own FP16 checkpoint at load time by bitsandbytes, so every operating point in this paper is produced by one algorithm on one codepath. A pre-quantized public checkpoint would introduce a different algorithm and different calibration data per model, which is a confound sitting directly on the variable being measured.

\subsection{Platform and Software}

Every measurement reported in this paper was taken on Modal, where each stage runs in its own serverless container holding a single A100 80GB. Containers are billed per second of execution and do not outlive their stage, so no measurement inherits a device warmed by the stage before it, and two persistent volumes carry the model cache and the results so that a failure late in a sequence never discards completed work. Table 2 records the software stack.

Two A100 variants were used, and the distinction is recorded here because measurements are split across them. The three per-model drift runs, the replay intervention and both kernel stages ran on the SXM4 part, quoted at 2039 GB/s of HBM2e bandwidth, 312 TFLOP/s of peak FP16 tensor throughput and a ridge point of 156 FLOP/byte. The 17-configuration sweep behind Tables 9, 10 and 12 ran on the PCIe part at 1935 GB/s. Where the two overlap they agree: INT4 measured on SXM4 in Table 8 and NF4 with double quantization measured on PCIe in Table 10 both give a Jaccard drift of 0.1142. The variant therefore does not move the drift figures reported here.

\begin{table}[t]
\caption{Software stack}
\label{tab:software}
\centering
\begin{tabular}{@{}ll@{}}
\toprule
\textbf{Component} & \textbf{Version} \\
\midrule
GPU & A100 80GB (SXM4, PCIe) \\
Platform & Modal (serverless) \\
PyTorch & 2.5.0+cu124 \\
CUDA / cuDNN & 12.4 / 9.1.0 \\
transformers & 4.46.0 \\
bitsandbytes & 0.44.1 \\
Triton & 3.1.0 \\
Python & 3.11.12 \\
\bottomrule
\end{tabular}
\end{table}

The transformers pin is load-bearing rather than incidental. From version 5 onward the library fuses MoE experts into packed three-dimensional parameters, and because bitsandbytes only substitutes nn.Linear modules, a nominally quantized MoE load silently quantizes attention alone. On Qwen3-30B-A3B that produced a report of zero replaced Linear layers for a 48-layer, 128-expert model. Every image is pinned below version 5 for this reason and the pipeline now hard-fails when an expert weight survives a quantized load unchanged.

\subsection{Sub-study 1: Triton Kernel Engineering}

Two fused Triton kernels were written. FusedRMSNorm \cite{ref6} merges variance computation and normalization into a single SRAM-resident pass, removing the redundant HBM traffic of the stock two-pass implementation. Stock PyTorch dispatches RMSNorm as three kernels, each reading and writing HBM in full; ours computes the squared mean, the normalization, and the weight scaling in one round trip, with a block size of 1024 for hidden dimensions up to 2048 and 2048 above that. FusedSoftmax applies the same treatment to the row-wise gate softmax using the online max-subtract-exp-sum formulation, collapsing four dispatches into one tile and writing HBM exactly once. Gate probabilities are accumulated in FP32 before the top-k, since the comparison that selects experts is precisely where a half-precision reduction order would change the answer.

Both kernels are installed by monkey-patching rather than by editing model source. RMSNorm modules are matched by class name and swapped; the router's softmax is intercepted by wrapping the MoE block's forward and dispatching only tensors shaped [batch, num\_experts] to the Triton path, leaving every other softmax untouched. Table 3 lists the five experiments.

\begin{table}[t]
\caption{Kernel experiment design}
\label{tab:kernel-design}
\centering
\small
\begin{tabularx}{\columnwidth}{@{}
  >{\raggedright\arraybackslash}p{1.12in}
  >{\raggedright\arraybackslash}X
  >{\raggedright\arraybackslash}X@{}}
\toprule
\textbf{Experiment} & \textbf{Configuration} & \textbf{Metric} \\
\midrule
1a: isolated RMSNorm & hidden 512--4096 & Median latency, GB/s \\
1b: isolated softmax & $N=64$ experts & Median latency, GB/s \\
1c: module fractions & 81 wrapped modules & \% of forward pass \\
1d: end to end & seq 128--1024, batch 1--4 & p50 latency, speedup \\
1e: compile sweep & 5 configurations & Speedup, graph breaks \\
\bottomrule
\end{tabularx}
\end{table}

Isolated benchmarks use 1000 warm-up and 5000 timed iterations and report the median. End-to-end runs use a fixed prompt padded to the target length, 50 warm-up and 200 timed forward passes, with peak memory from torch.cuda.max\_memory\_allocated. The Amdahl ceiling follows from the classical form, where f is the fraction of the forward pass spent in the optimized operations and s their isolated speedup:

\begin{equation}
S = \frac{1}{(1 - f) + f/s}
\end{equation}

How f is obtained is worth stating precisely. Summing the durations of CUDA kernels whose names match RMSNorm undercounts twice over: it misses the variance-reduction kernel, whose name does not match, and it misses the fact that QK-normalization gives each attention layer two further RMSNorm modules beyond the two in the residual stream. We instead wrap the actual modules in profiler record\_function ranges and attribute by range, covering 81 modules, which gives f $=$ 7.87\% and an Amdahl ceiling of 1.072x.

\subsection{Sub-study 2: Routing Drift and Its Causal Role}

Forward hooks were registered on every gate module, located by scanning named\_modules for gates whose parent is an MoE block. On each call the hook records the selected expert indices, keyed by layer and token position, and writes them to disk. One hundred MMLU-style prompts spanning a range of subjects were processed identically at every precision, giving 119,952 token positions per configuration. Table 4 defines the metrics computed against the FP16 reference.

\begin{table}[t]
\caption{Routing drift metrics}
\label{tab:metrics}
\centering
\small
\begin{tabularx}{\columnwidth}{@{}
  >{\raggedright\arraybackslash}p{0.88in}
  >{\raggedright\arraybackslash}X
  >{\raggedright\arraybackslash}X@{}}
\toprule
\textbf{Metric} & \textbf{Definition} & \textbf{Interpretation} \\
\midrule
Routing Similarity & Tokens with identical top-$k$ set & 1.0 $=$ perfect agreement \\
Jaccard Drift & $1 - |A\cap B|/|A\cup B|$ & 0.0 $=$ no drift \\
Overlap@$k$ & Mean $|A\cap B|/k$ & 1.0 $=$ same experts \\
Selection Shift & Fraction of slots changed & 0.0 $=$ no change \\
Swaps per token & $k\times$ Selection Shift & Comparable across $k$ \\
\bottomrule
\end{tabularx}
\end{table}

The last row of Table 4 exists because raw Jaccard drift is not comparable across models with different k. A single swapped expert out of k registers as 2/(k+1), which is 0.222 at top-8 but 0.286 at top-6, so a model with narrower selection appears to drift more for the same physical event. Multiplying selection shift by k gives the expected number of swapped experts per token, which is comparable, and doing so reverses the apparent INT4 ordering between DeepSeek and OLMoE.

Correlating drift with quality cannot establish that drift causes the quality loss, so the central experiment here is an intervention. The quantized model's expert selections are recorded, then the FP16 model is run again with its routing overridden to follow those recorded selections while every weight stays at full precision. If routing changes are what damage output, the replayed model should lose most of what the quantized model loses. A control replays FP16's own routes through the same machinery, which must return the baseline exactly; any deviation would be an artifact of the hooks rather than a result. Quality is reported as negative log-likelihood over all 119,952 positions rather than as task accuracy, because accuracy at a feasible evaluation budget cannot resolve the differences involved.

\subsection{Sub-study 3: Graph Breaks and Compilation}

Graph breaks were counted on the real 16-layer OLMoE checkpoint with torch.\_dynamo.explain, at the full hidden size so that the expert matrix multiplications are not artificially shrunk. Tracing does not execute kernels, so this stage needs no GPU and its result is device-independent.

Five configurations were then timed at identical shapes: eager as the reference, eager with the Triton kernels, torch.compile at default settings, torch.compile with capture\_dynamic\_output\_shape\_ops enabled, and that last configuration with the kernels also installed. The cuDNN attention backend is disabled throughout this sub-study, because Inductor's tensor layouts make it fail outright; timings within the sub-study are therefore internally consistent but are not directly comparable to the end-to-end numbers in Sub-study 1, which run with it enabled.

\section{Results}

\subsection{Triton Kernel Results}

\subsubsection{Isolated Performance}

Table 5 reports the isolated RMSNorm benchmark. Speedup rises with hidden dimension, from 5.62x at 512 to 8.98x at 4096, and the kernel reaches 1620 GB/s at the largest size, which is 79\% of the SXM4 part's 2039 GB/s against 9\% for the stock implementation. Fusing the two passes removes a third of the memory traffic, and since both versions are memory-bound the traffic saved is the speedup gained.

\begin{table}[t]
\caption{Isolated RMSNorm benchmark (A100)}
\label{tab:rmsnorm}
\centering
\begin{tabular}{@{}lcccc@{}}
\toprule
\textbf{Hidden} & \textbf{Baseline} & \textbf{Triton} & \textbf{Speedup} & \textbf{GB/s} \\
\midrule
512 & 0.0581 ms & 0.0103 ms & 5.62x & 609 \\
1024 & 0.0755 ms & 0.0132 ms & 5.72x & 953 \\
2048 & 0.1365 ms & 0.0188 ms & 7.27x & 1342 \\
4096 & 0.2789 ms & 0.0311 ms & 8.98x & 1620 \\
\bottomrule
\end{tabular}
\end{table}

The gate softmax behaves differently. It runs 0.0173 ms at baseline and 0.0086 ms fused, a 2.01x gain that does not improve with size because there is no size to grow: a 64-wide row dispatches a single warp with most lanes idle, and occupancy sits near 45\%. Both kernels have arithmetic intensities below 1.0 FLOP/byte, far under the 156 FLOP/byte ridge point, so neither is anywhere near compute-limited.

\subsubsection{Share of the Forward Pass}

An isolated speedup only matters in proportion to the time the operation occupies. Table 6 reports the forward-pass share of each optimized operation, attributed by wrapped module.

\begin{table}[t]
\caption{Forward-pass share and Amdahl ceiling}
\label{tab:fwdshare}
\centering
\begin{tabular}{@{}lc@{}}
\toprule
\textbf{Quantity} & \textbf{Value} \\
\midrule
RMSNorm share & 7.70\% \\
Router softmax share & 0.17\% \\
Combined f & 7.87\% \\
Effective kernel speedup & 6.88x \\
Amdahl ceiling & 1.072x \\
\bottomrule
\end{tabular}
\end{table}

The RMSNorm modules occupy 7.70\% of the forward pass and the router softmax 0.17\%, for a combined f of 7.87\% and an Amdahl ceiling of 1.072x. Even at this favourable ceiling, it is the end-to-end measurement that then refuses to deliver.

\subsubsection{End to End}

Table 7 gives the end-to-end sweep. At seq 128, batch 1, the patched model is 21.5\% slower: the kernels are launched more often than the work justifies and the launch cost dominates. The largest configuration reaches 1.027x. At seq 512, batch 4, which is the shape the module fractions in Table 6 were measured at and therefore the only row directly comparable to the 1.072x ceiling, the result is 0.999x.

\begin{table}[t]
\caption{End-to-end OLMoE latency, baseline versus Triton kernels}
\label{tab:endtoend}
\centering
\begin{tabular}{@{}ccccc@{}}
\toprule
\textbf{seq} & \textbf{batch} & \textbf{Baseline} & \textbf{Kernels} & \textbf{Speedup} \\
\midrule
128 & 1 & 246.8 ms & 314.5 ms & 0.785x \\
128 & 4 & 329.1 ms & 331.0 ms & 0.994x \\
512 & 1 & 319.8 ms & 322.0 ms & 0.993x \\
512 & 4 & 342.5 ms & 342.7 ms & 0.999x \\
1024 & 1 & 337.6 ms & 339.5 ms & 0.994x \\
1024 & 4 & 383.4 ms & 373.4 ms & 1.027x \\
\bottomrule
\end{tabular}
\end{table}

A ceiling of 1.072x and an achieved 0.999x at the same shape on the same device leaves a gap that Amdahl's law does not explain, because Amdahl already accounts for the 92\% of runtime the kernels do not touch. The explanation is in the baseline column of Table 7. Going from 128 tokens to 4096, a thirty-two-fold increase in work, raises latency from 246.8 ms to 383.4 ms, a factor of 1.55. A model whose latency is nearly independent of how much arithmetic it performs is not waiting on arithmetic. OLMoE's expert dispatch is a Python loop over 64 experts in each of 16 layers, on the order of a thousand small sequential kernel launches per forward pass, and a faster RMSNorm does not shorten a queue of launches. The kernels are integration-bound, not Amdahl-bound.

\subsection{Routing Drift Results}

\subsubsection{Overall Drift}

Table 8 reports drift for OLMoE at its native top-8, with intervals bootstrapped over prompts rather than token rows, since token rows within a prompt are not independent.

\begin{table}[t]
\caption{Routing drift, OLMoE-1B-7B at top-8, 119,952 token positions}
\label{tab:drift}
\centering
\begin{tabular}{@{}lcccc@{}}
\toprule
\textbf{Precision} & \textbf{Jaccard drift} & \textbf{95\% CI} & \textbf{Sel. shift} & \textbf{dNLL} \\
\midrule
INT8 & 0.0488 & [0.0477, 0.0501] & 0.0277 & +0.0031 \\
INT4 & 0.1142 & [0.1123, 0.1163] & 0.0660 & +0.0866 \\
\bottomrule
\end{tabular}
\end{table}

The routing is not stable. INT4 changes on average 0.53 of the eight selected experts per token position and INT8 0.22, and the two intervals are far from overlapping. It is worth being precise about what a Jaccard drift of 0.1142 means at top-8, because the figure sounds small: that 0.53 is a selection shift of 0.0660 across eight slots, so the typical affected token has one expert of eight replaced and a minority have two. The experts that change are overwhelmingly at the 8th and 9th ranked positions, where the logit margin is thinnest, and the top few ranks are essentially never disturbed at either precision. Repeated passes at a fixed precision are bit-identical, so none of this is sampling noise.

\begin{figure}[t]
\centering
\includegraphics[width=\columnwidth]{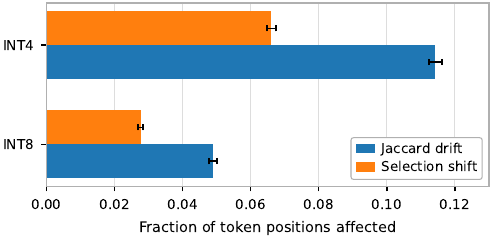}
\caption{Jaccard drift and selection shift for INT8 and INT4 on OLMoE, with 95\% intervals bootstrapped over prompts rather than token rows. INT4 drift is 2.3x INT8 and the intervals are disjoint.}
\label{fig:1}
\end{figure}

\subsubsection{Layer Coverage}

Quantizing only the first N of 16 layers turns coverage into a dial, and Table 9 shows drift increasing monotonically along it while quality saturates early.

\begin{table}[t]
\caption{Layer coverage dial, NF4 on the first N of 16 layers}
\label{tab:layercov}
\centering
\begin{tabular}{@{}lccccc@{}}
\toprule
\textbf{Layers quantized} & \textbf{2} & \textbf{4} & \textbf{8} & \textbf{12} & \textbf{16} \\
\midrule
Jaccard drift & 0.0626 & 0.0814 & 0.0978 & 0.1074 & 0.1142 \\
dNLL & +0.0168 & +0.0791 & +0.0835 & +0.0836 & +0.0866 \\
\bottomrule
\end{tabular}
\end{table}

Drift keeps climbing from 0.0626 at two layers to 0.1142 at sixteen, but almost all of the NLL penalty has already been paid by layer four. The two quantities therefore do not move together along this dial, which is the first sign that drift is not the mechanism of the damage.

\begin{figure}[t]
\centering
\includegraphics[width=\columnwidth]{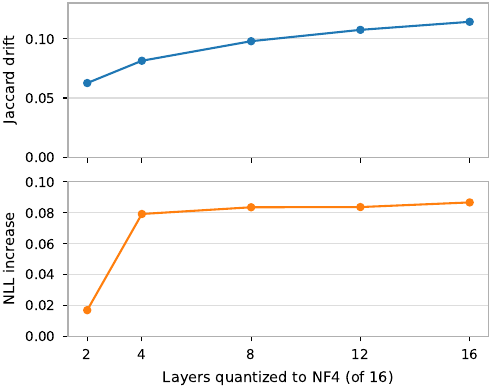}
\caption{NF4 applied to the first N of 16 layers. Drift (top) rises monotonically with coverage while the NLL penalty (bottom) has almost all been paid by layer four, so the two do not move together along this dial.}
\label{fig:2}
\end{figure}

\subsubsection{Which Knobs Matter}

Table 10 sorts the quantization configurations by drift, which makes it clear that one knob dominates and the rest are close to irrelevant.

\begin{table}[t]
\caption{Quantization knobs, sorted by drift}
\label{tab:knobs}
\centering
\begin{tabular}{@{}llcc@{}}
\toprule
\textbf{Configuration} & \textbf{Parameter} & \textbf{Jaccard drift} & \textbf{dNLL} \\
\midrule
int8\_t3 & outlier thr. 3.0 & 0.0476 & +0.0050 \\
int8\_t6 & outlier thr. 6.0 (default) & 0.0488 & +0.0031 \\
int8\_t12 & outlier thr. 12.0 & 0.0498 & +0.0264 \\
int8\_t0 & thr. 0.0, decomp. off & 0.0517 & +0.0535 \\
nf4 & NF4 baseline & 0.1140 & +0.0872 \\
nf4\_fp32c & FP32 compute & 0.1140 & +0.0871 \\
nf4\_dq & double quant. & 0.1142 & +0.0866 \\
fp4 & FP4 data type & 0.1524 & +0.1135 \\
fp4\_dq & FP4 + double quant. & 0.1526 & +0.1126 \\
\bottomrule
\end{tabular}
\end{table}

The 4-bit data type is what matters: FP4 drifts 34\% more than NF4 at identical bit width, reflecting NF4's better fit to normally distributed weights. Double quantization and FP32 compute change drift in the fourth decimal place. The INT8 outlier threshold is the more instructive lever, and it needs stating precisely because one of its settings is not a value on the same axis as the others. The threshold is the magnitude above which an activation column is held in FP16 instead of being quantized, so raising it from 3.0 to 12.0 protects progressively fewer columns, while setting it to 0.0 disables the mixed-precision decomposition altogether and quantizes every column. Ordered by that aggressiveness the drift is monotone across the four settings, 0.0476 to 0.0517, a span of 8\%. Over the same four settings dNLL rises more than tenfold, from +0.0050 to +0.0535, and not monotonically: the library default of 6.0 gives the lowest loss of the four. A lever that perturbs the activations arriving at the gate, moves routing by 8\%, and moves output quality by an order of magnitude is hard to reconcile with routing being the mechanism of the damage.

\subsubsection{Does Drift Cause the Damage?}

The intervention answers that directly. Table 11 gives the four NLL values.

\begin{table}[t]
\caption{Causal route replay}
\label{tab:replay}
\centering
\begin{tabular}{@{}lc@{}}
\toprule
\textbf{Configuration} & \textbf{NLL} \\
\midrule
FP16 baseline & 2.276696 \\
FP16 weights, FP16 routes (control) & 2.276696 \\
FP16 weights, INT4 routes & 2.279028 \\
INT4 & 2.363338 \\
\bottomrule
\end{tabular}
\end{table}

The control reproduces the FP16 baseline to all printed digits, so the replay machinery adds nothing of its own. Forcing full-precision weights to follow INT4's expert choices costs +0.002332 of NLL, against INT4's own +0.086642. Routing therefore accounts for 2.7\% of the degradation and weight error for the remaining 97.3\%. The experts are substitutable: sending a token to its 9th-best expert instead of its 8th-best costs very little, because at that point in the ranking the two experts do close to the same thing.

\begin{figure}[t]
\centering
\includegraphics[width=\columnwidth]{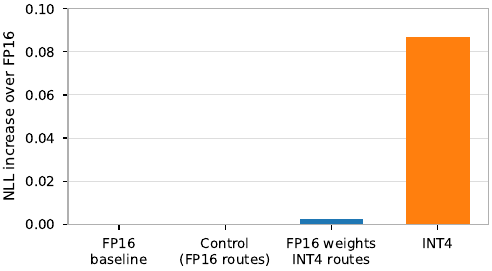}
\caption{Causal attribution, plotted as NLL increase over the FP16 baseline. The control returns exactly zero, confirming the replay machinery is neutral; replaying INT4's routes through FP16 weights recovers 2.7\% of INT4's increase, leaving 97.3\% to weight error.}
\label{fig:3}
\end{figure}

This is the result that reframes the sub-study. Drift is real but almost entirely harmless on its own.

\subsubsection{Drift, Gate Divergence, and Quality}

Across 16 distinct configurations drift does predict quality loss, with Pearson +0.907 and Spearman +0.891 against dNLL. Taken alone that would make a reasonable case for routing fidelity as a systems metric. The difficulty appears when gate KL divergence is fitted to the same target, giving +0.855 and +0.900: a continuous measure of how far the gate distribution moved predicts quality about as well as the discrete count of how many experts changed. The two predictors are 98\% collinear with one another (Pearson +0.978), so the sweep cannot separate them.

\begin{figure}[t]
\centering
\includegraphics[width=\columnwidth]{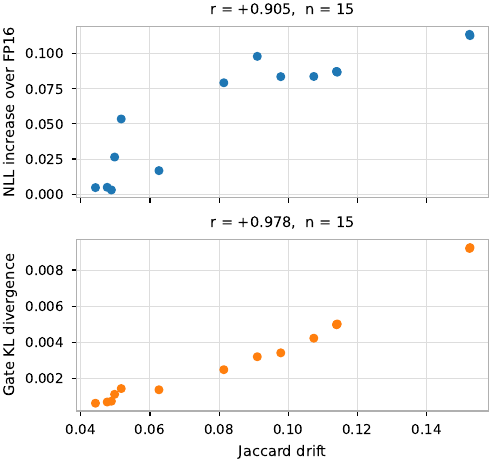}
\caption{Top: Jaccard drift against NLL increase across 16 distinct configurations, Pearson +0.907. Bottom: the same drift against gate KL divergence, Pearson +0.978, which is why correlation alone cannot separate the two predictors.}
\label{fig:4}
\end{figure}

A correlation of +0.907 against a substitute that is 98\% collinear and an intervention attributing 2.7\% of the effect are not in conflict; they describe a variable that tracks the damage without causing it. Drift is a symptom of gate perturbation, and gate perturbation accompanies the weight error that does the damage.

\subsubsection{Router Exemption}

If routing fidelity were the objective, protecting the routers should help. Table 12 tests that by quantizing everything except the 16 router modules.

\begin{table}[t]
\caption{Router exemption}
\label{tab:router-exempt}
\centering
\begin{tabular}{@{}llcc@{}}
\toprule
\textbf{Configuration} & \textbf{Routers} & \textbf{Jaccard drift} & \textbf{dNLL} \\
\midrule
int8\_t6 & quantized & 0.0488 & +0.0031 \\
int8\_gate\_fp16 & FP16 & 0.0442 & +0.0048 \\
nf4 & quantized & 0.1140 & +0.0872 \\
nf4\_gate\_fp16 & FP16 & 0.0910 & +0.0979 \\
\bottomrule
\end{tabular}
\end{table}

Drift falls by 20\% at NF4, from 0.1140 to 0.0910, and NLL gets worse, from +0.0872 to +0.0979. The same inversion appears at INT8. Two things follow. Mechanically, the router's own quantization causes about a fifth of the drift and upstream perturbation the other four fifths, since exempting the gate removes only that fifth. Practically, routing fidelity and output quality are separable objectives that can be traded against each other in the wrong direction, which is a stronger statement than saying they are weakly related.

\begin{figure}[t]
\centering
\includegraphics[width=\columnwidth]{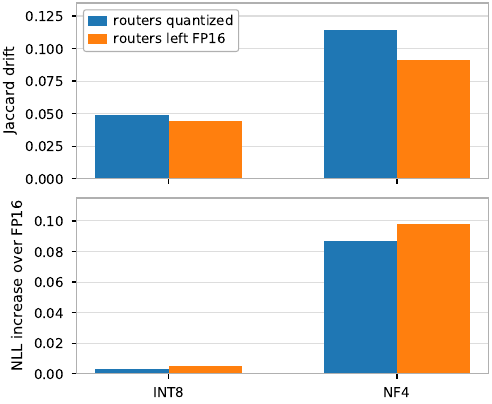}
\caption{Router exemption. Quantizing everything except the 16 router modules lowers Jaccard drift (top) while raising NLL (bottom), at both precisions, so routing fidelity and output quality move in opposite directions.}
\label{fig:5}
\end{figure}

\subsubsection{Across Architectures}

Table 13 extends the measurement to DeepSeek-V2-Lite and Qwen3-30B-A3B, reporting both raw Jaccard and the top-k-corrected swap count.

\begin{table*}[t]
\caption{Cross-architecture drift, raw and top-k corrected}
\label{tab:crossarch}
\centering
\begin{tabular}{@{}lccccccc@{}}
\toprule
\textbf{Model} & \textbf{k} & \textbf{Prec} & \textbf{Jaccard} & \textbf{Swaps/token} & \textbf{95\% CI} & \textbf{dNLL} \\
\midrule
DeepSeek-V2-Lite & 6 & INT8 & 0.0419 & 0.1475 & [0.1429, 0.1529] & +0.00118 \\
OLMoE-1B-7B & 8 & INT8 & 0.0488 & 0.2214 & [0.2160, 0.2273] & +0.00312 \\
Qwen3-30B-A3B & 8 & INT8 & 0.0690 & 0.3171 & [0.3063, 0.3290] & +0.00512 \\
DeepSeek-V2-Lite & 6 & INT4 & 0.1303 & 0.4686 & [0.4565, 0.4830] & +0.02524 \\
OLMoE-1B-7B & 8 & INT4 & 0.1142 & 0.5281 & [0.5189, 0.5384] & +0.08664 \\
Qwen3-30B-A3B & 8 & INT4 & 0.1657 & 0.7961 & [0.7792, 0.8150] & +0.05810 \\
\bottomrule
\end{tabular}
\end{table*}

All six intervals within a precision are disjoint. The corrected column changes the reading of the INT4 rows: DeepSeek has the higher raw Jaccard, 0.1303 against OLMoE's 0.1142, but swaps fewer experts per token, 0.4686 against 0.5281, because its top-6 selection inflates the same physical event. Any cross-model claim has to use the corrected quantity. The pattern that survives is that finer selection granularity drifts more, with Qwen's 128 experts at top-8 drifting most on every measure.

\begin{figure}[t]
\centering
\includegraphics[width=\columnwidth]{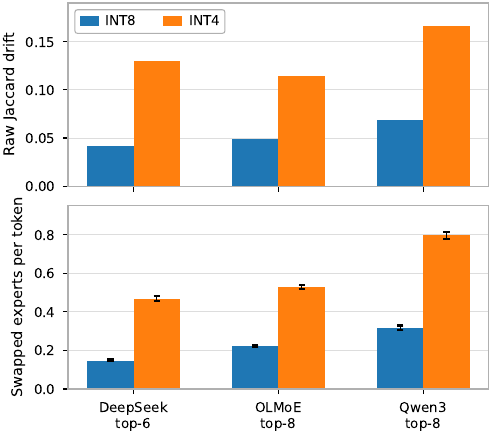}
\caption{Raw Jaccard drift (top) against expected swapped experts per token (bottom). Correcting for top-k reverses the INT4 ordering between DeepSeek-V2-Lite and OLMoE, since top-6 selection inflates the same physical event.}
\label{fig:6}
\end{figure}

Drift ranks the three models by quality loss perfectly at INT8, Spearman +1.00, but only partly at INT4, +0.50. Three points establish an ordering rather than a fit, and capacity is not controlled: the INT4 disagreement falls precisely on the 7B-versus-30B pair. DeepSeek's unquantized gate is a further confound, and it is a plausible mechanism for its advantage as well as a reason not to read that advantage as architectural.

\subsubsection{Task Accuracy}

Table 14 gives task accuracy for completeness, and explains why the rest of this paper reports NLL instead.

\begin{table}[t]
\caption{Task accuracy at lm\_eval\_limit 500}
\label{tab:accuracy}
\centering
\begin{tabular}{@{}lcc@{}}
\toprule
\textbf{Precision} & \textbf{MMLU} & \textbf{HellaSwag} \\
\midrule
FP16 & 0.5430 & 0.7060 \\
INT8 & 0.5419 & 0.7020 \\
INT4 & 0.5320 & 0.6840 \\
\bottomrule
\end{tabular}
\end{table}

No drop in Table 14 is distinguishable from zero at the evaluation budget used; the largest is about 0.8 sigma. NLL over 119,952 token positions has far lower variance than a few hundred multiple-choice outcomes, which is why it carries the causal result.

\subsection{torch.compile Results}

\subsubsection{Graph Breaks}

Tracing the real 16-layer checkpoint finds 23 graph breaks. Sixteen of the 23 occur at one site, a torch.nonzero in the expert dispatch whose output shape depends on the data, and the remainder are conditionals on CUDA tensors nearby. Setting torch.\_dynamo.config.capture\_dynamic\_output\_shape\_ops removes all of them, which makes the break count fully controllable and turns the interesting question into what happens when it reaches zero.

\subsubsection{Compile Benchmark}

Table 15 times five configurations at two identical shapes.

\begin{table}[t]
\caption{Compile benchmark, five configurations at identical shapes}
\label{tab:compile}
\centering
\begin{tabular}{@{}lcccc@{}}
\toprule
\textbf{Configuration} & \textbf{512x4} & \textbf{1024x4} & \textbf{Breaks} & \textbf{First fwd} \\
\midrule
eager & 1.000x & 1.000x & 0 & 1.0 s \\
eager + kernels & 0.979x & 1.033x & 0 & 2.8 s \\
compile, default & 0.822x & 0.878x & 19 / 36 & 30--61 s \\
compile + capture & 0.613x & 0.327x & 0 & 37--40 min \\
\bottomrule
\end{tabular}
\end{table}

Compiling at default settings costs 18\% at seq 512. Removing every graph break costs 39\% at that shape and 67\% at seq 1024, and raises the first forward pass from one second to nearly forty minutes. Zero graph breaks is three times slower than eager, which is the clearest result in this paper and the one most at odds with how break counts are usually reported. Unbacked symbolic shapes force Inductor to generate code that handles any size, and the guards and fallbacks that entails cost more than the eager dispatch they replace.

\begin{figure}[t]
\centering
\includegraphics[width=\columnwidth]{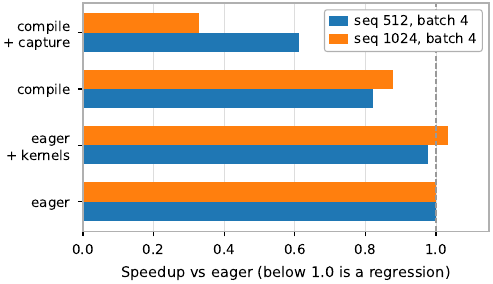}
\caption{Speedup against eager at two shapes, with graph-break counts annotated. Compiling costs 18\% at seq 512; removing every break with dynamic shape capture costs 39\% there and 67\% at seq 1024.}
\label{fig:7}
\end{figure}

The fifth configuration was the hypothesis this benchmark existed to test: that fusing the dispatch would amortize the launches and finally let the Triton kernels pay. Since fusing the dispatch is itself a threefold loss, the premise fails before the kernels get a chance, and the combination is not worth reporting separately.

\section{Conclusion}

Three optimizations, measured on three models, and none of them pays. The reasons are more useful than the fact, and they reduce to two mechanisms.

The model is launch-bound. Thirty-two times the tokens cost 1.55 times the time, which means the device is waiting on a thousand small sequential kernel launches rather than on arithmetic. That single fact explains why 5.6x to 9.0x isolated kernels return 0.999x against a real 1.072x ceiling, and why compilation cannot rescue them: the compiler's own attempt to fuse the dispatch is a threefold regression.

The experts are substitutable. INT4 changes on average 0.53 of the eight selected experts per token, but replaying those exact changes through full-precision weights reproduces 2.7\% of the loss. Drift correlates with quality at +0.907 while being 98\% collinear with gate divergence, so the correlation was never evidence of a mechanism. Exempting the routers from quantization sharpens the point by reducing drift 20\% and making the model worse.

Taken together these say something specific about where MoE inference effort belongs. Protecting routing fidelity is not it, and neither is reducing graph-break counts. The binding constraint is the launch structure of the expert dispatch, and the optimization that would matter is the one that makes the dispatch a single batched operation rather than a loop. Not because that would unlock the kernel gains, which we now know it would not on its own, but because it is the only one of these three levers attached to the quantity that actually sets the latency.

Limitations: The causal replay has run on OLMoE alone, so the 2.7\% attribution is a single-model result. The correlation rests on 16 configurations derived from one checkpoint and one prompt set, which measures how drift and quality move together across quantization settings rather than across models or corpora. Drift and gate KL cannot be separated by correlation at all, given 98\% collinearity; only the intervention distinguishes them. The cross-model ordering rests on three points with capacity uncontrolled and DeepSeek's gate unquantized. Compiler timings run with cuDNN attention disabled and so are not directly comparable to the end-to-end table. Task accuracy differences sit below the noise floor at the budget used. Finally, all three models are quantized by a single library, so the results characterize how routing responds to that quantizer rather than to 4-bit quantization in general.

Future work: Three directions follow directly. Hand-quantizing DeepSeek's nn.Parameter gate would confirm or eliminate the confound behind its apparent advantage, and it is a one-stage experiment. Batching the expert dispatch into a single grouped matrix multiplication attacks the launch-bound structure that Sub-study 1 identifies as the real constraint, and unlike a custom Dynamo lowering rule it does not depend on symbolic shapes being cheap. Finally, a quantization objective penalizing routing flips would now be worth testing against the router-exemption result rather than against drift alone, since we have shown that lowering drift can raise loss.

\section*{Code Availability}

The complete source code, the Modal execution harness, and the raw per-token route dumps from which every metric in this paper recomputes are open source and available at \url{https://github.com/GokuHashira/moe-ceilings}.

\section*{Acknowledgements}

The authors thank the University of Maryland for access to the Zaratan high-performance computing cluster, which supported exploratory experimentation during the initial phases of this work, and Modal for the serverless GPU capacity on which every result reported here was produced: an NVIDIA A100 80GB SXM4 for the three per-model drift runs, the replay intervention and both kernel stages, and an NVIDIA A100 80GB PCIe for the 17-configuration quantization sweep.

\end{document}